\documentclass[%
 reprint,
superscriptaddress,
 amsmath,amssymb,
 aps,prb,
]{revtex4-2}

\usepackage{amsmath}
\usepackage{xcolor}
\usepackage{makecell}
\usepackage[mathscr]{euscript}
\usepackage{graphicx}
\usepackage{dcolumn}
\usepackage{bm}
\usepackage{hyperref}

\usepackage{tabularx,ragged2e}
\usepackage{multirow}

\newcolumntype{C}{>{\Centering\arraybackslash}X} 

\begin{document}



\title{Comparative Assessment of Thermal Transport Theories: Dual-Channel Mechanism Dictates Heat Transport in Ultralow-\texorpdfstring{$\kappa$}{kappa} Materials}

\author{Soham Mandal}
\affiliation{Centre for Condensed Matter Theory, Department of Physics, Indian Institute of Science, Bangalore 560012, India}
\author{Ashutosh Srivastava}
\affiliation{Materials Research Centre, Indian Institute of Science, Bangalore 560012, India}
\author{Tanmoy Das}%
\affiliation{Centre for Condensed Matter Theory, Department of Physics, Indian Institute of Science, Bangalore 560012, India}
\author{Manish Jain}%
\affiliation{Centre for Condensed Matter Theory, Department of Physics, Indian Institute of Science, Bangalore 560012, India}
\author{Abhishek Kumar Singh}
\affiliation{Materials Research Centre, Indian Institute of Science, Bangalore 560012, India}
\author{Prabal K. Maiti}
\email{maiti@iisc.ac.in}
\affiliation{Centre for Condensed Matter Theory, Department of Physics, Indian Institute of Science, Bangalore 560012, India}

\date{\today}

\begin{abstract}
Anomalous heat transport in strongly anharmonic crystalline solids poses both a fundamental challenge to the theoretical understanding and an opportunity for thermoelectric and thermal barrier coating applications. 
Although the Green-Kubo framework accurately reproduces experimental thermal conductivities ($\kappa$) at high temperatures, it provides limited microscopic insight, neglects the Bose-Einstein statistics of lattice vibrations, and remains computationally demanding when implemented within first-principles atomistic simulations.
On the other hand, the conventional Boltzmann transport equation (BTE) fails in these materials, as strong anharmonicity drives phonons into an overdamped regime, invalidating the phonon-gas picture.
Herein, the thermal transport properties in TlAgSe, a metal chalcogenide, and Cs\textsubscript{2}PbI\textsubscript{2}Cl\textsubscript{2}, an all-inorganic layered Ruddlesden-Popper perovskite, are investigated by explicitly accounting for temperature-dependent lattice dynamics through machine learning interatomic potentials and employing the Wigner transport equation (WTE) framework.
Crucially, we find that heat conduction is governed not only by higher-order phonon scattering-dominated populations' transport channel described within the BTE, but significant contributions also arise from a coherences' channel in the WTE framework arising from wave-like interbranch coherence between eigenstates.
Incorporating four-phonon scattering, WTE predicts average room-temperature $\kappa$ values of 0.31 Wm\textsuperscript{-1}K\textsuperscript{-1} (TlAgSe) and 0.38 Wm\textsuperscript{-1}K\textsuperscript{-1} (Cs\textsubscript{2}PbI\textsubscript{2}Cl\textsubscript{2}), in excellent agreement with experiments and Green-Kubo predictions.
Phonon scattering-rate analysis reveals strong coherences' contributions and prevalent overdamped phonon modes, demonstrating the breakdown of the conventional BTE framework based on the phonon quasiparticle picture with only first-order anharmonic perturbation.
This computational approach provides a unified description of heat transport in ultralow-$\kappa$ materials, offering a basis for the rational design of phononic and thermoelectric devices.
\end{abstract}

\maketitle


\section{\label{sec:Introduction} INTRODUCTION}
Heat transport is a fundamental property of materials with far-reaching implications for both basic science and technological applications. In crystalline semiconductors and insulators, energy is predominantly carried by quantized normal modes of lattice vibrations, known as phonons. 
Consequently, understanding phonon-mediated transport and the associated emergent material properties has remained a central focus of condensed matter physics and materials science research over the past several decades.
Crystalline materials exhibiting intrinsically low and ultralow lattice thermal conductivity ($\kappa$) are of significant interest for a wide range of technological applications, particularly in thermoelectric energy conversion~\cite{TE_rev2_2008, TE_rev1_2022} and thermal barrier coatings~\cite{TBC1_2009, TBC2_2012}.
A key strategy to improve thermoelectric performance is to selectively scatter phonons, thereby suppressing $\kappa$ to an ultralow value while maintaining favorable electronic transport.\cite{Mukherjee2022}
The physical mechanism of heat transport in ultralow-$\kappa$ materials remains intricate, often exhibiting unconventional phenomena such as the breakdown of the phonon-gas model~\cite{Breakdown_phGas_2024}, a glass-like rise in temperature dependence in $\kappa$ in crystals~\cite{Glass_kappa_2021, Glass_kappa_2022}, anomalously strong lattice anharmonicity~\cite{Anharmonicity_Nat_2014, Anharmonicity_prb_24}, and ``rattling" mechanism~\cite{rattling_Nat_2008, Cs2SnI6_2022}.
Accordingly, theoretical modeling of thermal transport in ultralow-$\kappa$ materials demands careful reassessment of the applicability of conventional first-principles frameworks. 
The standard Peierls-Boltzmann transport equation (BTE)~\cite{PBTE_Peierls_1929}, based on the phonon quasiparticle picture and the phonon-gas model of interactions in crystalline solids, proves inadequate in capturing the experimentally observed behavior~\cite{CsBX3_pnas_2017}. 
Notably, this discrepancy persists even upon inclusion of computationally demanding four-phonon~\cite{Coherent_k_AJain_2020_Tl3VSe4} and higher-order scattering processes~\cite{FourPhonon_PRB_16, 5ph_6Ph_2025}.
To address this discrepancy, a unified theoretical framework based on the Wigner formulation of thermal transport (WTE) in solids has been developed~\cite{ TwoChannelTransport_2019, WignerTransport_PRX_2022}. 
Within this approach, it is proposed that, in addition to the particle-like propagation of phonon quasiparticles, heat carriers in such complex crystals also exhibit coherence between neighboring eigenstates, which transfer energy to one another without any of them propagating individually. The resulting transport is diffusive and closely analogous to that of glasses~\cite{AllenFeldman_PRL_1989}.
Although the WTE framework has proven effective in addressing discrepancies between experimental observations and BTE predictions~\cite{TwoChannelTransport_science_2018}, it relies on computationally demanding calculations of perturbative phonon scattering rates in such structurally complex materials.
Moreover, the strong anharmonicity in the interatomic interaction of these materials raises the question whether it can be treated as a ``small" perturbation.
As such, a framework capable of addressing arbitrary anharmonic strength that can bypass computationally demanding perturbative phonon-scattering rate calculations is especially sought after.
The determination of thermal transport coefficients through the Green-Kubo (GK) theory in the linear response regime of the system's heat current provides a compelling alternative framework to address these challenges~\cite{aiGK_natphy_2016, aiGK_prl_2017, aiGK_prl_2023, GK_CALF-20_25}. 
By leveraging equilibrium molecular dynamics (MD) simulations to compute the heat current, the nonperturbative GK framework inherently captures anharmonic phonon interactions to all orders within the underlying interatomic potential, as well as finite-temperature effects.
Besides the GK framework, the direct heating-cooling method in nonequilibrium MD~\cite{PCCP} and the homogeneous nonequilibrium MD (HNEMD) method~\cite{HNEMD_PhysLettA_1982} also incorporate full temperature-dependent dynamics and anharmonicity of all orders.
However, because of their foundation in classical MD, these frameworks do not incorporate Bose-Einstein (BE) statistics of phonons and nuclear quantum effects, which become important only at lower temperatures.
\par
Materials situated at the extreme limits of anharmonicity and exhibiting ultralow-$\kappa$ provide an ideal platform for critically assessing the strengths and limitations of competing theoretical frameworks for heat transport.
Two broad classes of materials exhibiting ultralow-$\kappa$, favorable for thermoelectric energy conversion, are metal chalcogenides and halide perovskites. 
Numerous compounds within these families of materials have been extensively investigated for modulating phonon-mediated heat transport, both experimentally~\cite{GeSe_2025, AgSbTe2_2025, CsBX3_pnas_2017, Cs2SnI6_2022} and theoretically~\cite{sml_2026, Tl2VSe4_PRL_2020_coh_trans, Cs2AgBiBr6_2024_npjComp}. 
Among them, TlAgSe, a metal chalcogenide, and Cs\textsubscript{2}PbI\textsubscript{2}Cl\textsubscript{2}, an all-inorganic layered Ruddlesden-Popper perovskite, have recently emerged as particularly promising candidates for thermoelectric applications, driven by their distinctive anharmonicity-driven structural characteristics and intrinsically suppressed lattice thermal transport.
A recent experimental study reports that TlAgSe possesses an ultralow thermal conductivity of $\sim$ 0.3 Wm\textsuperscript{-1}K\textsuperscript{-1} at room temperature~\cite{TlAgSe_exp}. 
This phenomenon is driven by lattice instability arising from cationic Ag-Ag repulsion within edge-sharing [AgSe\textsubscript{4}]\textsuperscript{-} tetrahedra (see inset of Fig.~\ref{fig:BTE_WTE_GK}c), which induces strong lattice anharmonicity and scatters phonons, thereby suppressing $\kappa$.
Experimental studies on Cs\textsubscript{2}PbI\textsubscript{2}Cl\textsubscript{2} reveal anisotropic thermal transport, with room-temperature $\kappa$ in the range of $\sim$ 0.37-0.41 Wm\textsuperscript{-1}K\textsuperscript{-1}~\cite{Cs2PbI2Cl2_k-2020}.
The presence of low-energy optical phonon modes leads to strong scattering of heat-carrying acoustic phonons, resulting in intrinsically ultralow $\kappa$ in Cs\textsubscript{2}PbI\textsubscript{2}Cl\textsubscript{2}.
Furthermore, a recent experimental study also reveals a distinct temperature-induced anharmonic lattice dynamics, resulting in local symmetry breaking and further contributing to the ultralow $\kappa$ in Cs\textsubscript{2}PbI\textsubscript{2}Cl\textsubscript{2}~\cite{Cs2PbI2Cl2-2024}.
A rigorous treatment of finite-temperature anharmonic lattice dynamics and higher-order phonon interactions is therefore essential for a comprehensive theoretical understanding of heat transport in these materials.
Conventional perturbative approaches typically rely on atomic displacements about a single stable equilibrium structure at 0 K to evaluate interatomic force constants (IFCs) and the resulting thermal transport coefficients~\cite{ShengBTE_14, Phono3py_FD}.
Such approximations are inadequate for such strongly anharmonic systems, where finite-temperature effects induce significant structural distortions and access to metastable configurations~\cite{Instability_perovskite_PRXE_24}.
However, a rigorous and accurate treatment of full lattice dynamics is, in principle, attainable only through computationally demanding {\em ab initio} MD based on density functional theory (DFT). 
For such structurally complex materials containing heavy atoms, the prohibitive computational cost makes the direct application of {\em ab initio} MD largely impractical. 
The recent advancement of machine learning interatomic potential (MLIP)~\cite{MLIP, DeePMD_JCP_2023} in the field of molecular simulations offers an alternative route, which bypasses the computational cost of DFT while retaining a similar level of accuracy.
Our recent theoretical study~\cite{sml_2026} leveraging MLIP-based MD and GK framework for the heat transport has successfully captured the finite-temperature structural distortions and ultralow $\kappa$ of TlAgSe and Cs\textsubscript{2}PbI\textsubscript{2}Cl\textsubscript{2} observed in experiments.
Although atomic vibrations can be explicitly observed in MD simulations and the GK formalism yields reliable $\kappa$ values, extracting detailed microscopic insights into phonon transport mechanisms from such a nonperturbative approach alone remains nontrivial. 
Consequently, for the rational design of next-generation ultralow-$\kappa$ thermoelectric materials, it is essential to interpret the underlying physical origins of their unconventional transport properties.
In this context, a comparative analysis of different theoretical frameworks for thermal transport in these materials that incorporate hierarchical levels of anharmonic phonon interactions and account for full temperature-dependent lattice dynamics remains lacking in the current literature.
\par
In this work, we investigate the thermal transport properties of TlAgSe and Cs\textsubscript{2}PbI\textsubscript{2}Cl\textsubscript{2} using MLIP-based MD simulations, and uncover the underlying physical mechanisms governing their ultralow $\kappa$ through a detailed analysis of perturbative BTE and WTE formalisms, with direct comparison to the GK predictions and experimental observations.
\par
The rest of the paper is organized as follows: We first briefly discuss the WTE formalism, highlighting the additional coherences' contribution ($\kappa^\mathrm{C}$) beyond the conventional populations' term ($\kappa^\mathrm{P}$) described within the BTE framework.  
We then present the temperature-dependent $\kappa$ of both materials, decomposed into $\kappa^\mathrm{C}$ and $\kappa^\mathrm{P}$, and benchmark the results against Green-Kubo predictions and experimental data. 
Subsequently, we examine the role of higher-order phonon interactions, demonstrating that the inclusion of four-phonon scattering processes is essential for the applicability of perturbative approaches in such strongly anharmonic systems and for achieving agreement with experiment. 
Finally, we analyze the distinct heat transport regimes governed by the magnitude of phonon scattering rates, assess the validity of different theoretical frameworks across these regimes, and summarize our key findings.

\section{\label{sec:Methods} METHODS}
Finite-temperature lattice dynamics were integrated into the perturbative thermal transport frameworks by performing MD simulations, using MLIPs, for both the materials.
The \texttt{DeePMD-kit} software package~\cite{DeePMD_JCP_2023} was employed to train artificial neural network-based MLIPs for both the TlAgSe and Cs\textsubscript{2}PbI\textsubscript{2}Cl\textsubscript{2} systems.
Details of the {\em ab initio} MD simulations used to generate the training dataset, along with the DeePMD training parameters and associated loss metrics, were provided in our previous work~\cite{sml_2026}. 
IFCs were extracted from the MD trajectories employing our in-house developed \texttt{SPRING} code~\cite{arXiv, SPRING_GitHub}, which was subsequently used to compute $\kappa$ via full iterative solutions of the BTE and WTE.
Classical MD simulations using the MLIPs were performed employing the \texttt{LAMMPS} package~\cite{LAMMPS} on a 7$\times$7$\times$7 supercell of both the materials.
The integration timestep was set to 1 fs. 
Each system was initially equilibrated for 2 ns NPT simulation, followed by a 500 ps NVT run to achieve thermalization at the target temperature. 
Subsequently, a 2 ns production simulation was carried out in the microcanonical (NVE) ensemble.
For both TlAgSe and Cs\textsubscript{2}PbI\textsubscript{2}Cl\textsubscript{2}, 100 snapshots were selected at 20 ps intervals from the final 2 ns NVE simulation to construct the force-displacement dataset.
IFCs up to fourth order were obtained by performing a least-squares fitting on the force-displacement datasets at each temperature. 
Long-range forces arising from dipole-dipole interactions were separated out by calculating the long-range second-order IFCs from Born effective charges and dielectric tensors of the materials~\cite{EwaldNonAnalytic2}. 
The Born effective charges and dielectric tensors of both materials were computed using density functional perturbation theory (DFPT) implemented in the \texttt{VASP} package~\cite{VASP}.
For a detailed description of the IFC calculation methodology, readers are referred to our previous work~\cite{arXiv}.
The phonon scattering rates ($\tau^{-1}$) calculation includes contributions from three-phonon processes mediated by third-order anharmonic IFCs, four-phonon processes governed by fourth-order anharmonic IFCs, and phonon-isotope impurity scattering.
Energy conservation among the interacting phonon modes ($\boldsymbol{q}s$) in Fermi's golden rule is implemented using a parameter-free tetrahedron integration over the Brillouin zone (BZ).
Convergence of the $\kappa$ values from the BZ integration is achieved using a 17$\times$17$\times$17 $\mathbf{q}$-point mesh for both the three-phonon and four-phonon scattering processes.
The detailed expressions for different terms in $\tau^{-1}$ can be found in Ref.~\cite{arXiv}
After obtaining $\tau^{-1}$, we calculated the thermal conductivity tensor for both the particle-like propagation (populations' contribution, $\kappa^{\mathrm{P}}$) of phonon-quasiparticles using BTE and the interbranch coherence (coherences' contribution, $\kappa^{\mathrm{\mathrm{C}}}$) derived from the WTE.
The expressions for $\kappa^{\mathrm{P}}$ and $\kappa^{\mathrm{C}}$ can be combined into a single unified equation as follows~\cite{WignerTransport_PRX_2022}:
\begin{widetext}
\begin{equation}\label{eq:kappaP_C}
\begin{aligned}
\kappa_{\alpha \beta}^{\mathrm{P/C}} = \frac{\hbar^2}{k_B T^2NV} \sum\limits_{\boldsymbol{q}}\sum\limits_{s,s^{'}} \frac{\omega(\boldsymbol{q}s) + \omega(\boldsymbol{q}s^{'})}{2} v_{\boldsymbol{q}, ss^{'}}^{\alpha}  v_{\boldsymbol{q}, s^{'}s}^{\beta}  \frac{\omega(\boldsymbol{q}s) \frac{n(\boldsymbol{q}s) (n(\boldsymbol{q}s)+1)}{2} + \omega (\boldsymbol{q}s^{'}) \frac{n(\boldsymbol{q}s^{'}) (n(\boldsymbol{q}s^{'})+1)}{2}}{(\omega(\boldsymbol{q}s) - \omega(\boldsymbol{q}s^{'}))^2 + \left(\frac{\Gamma(\boldsymbol{q}s) + \Gamma(\boldsymbol{q}s^{'})}{2}\right)^2} \; \frac{\Gamma(\boldsymbol{q}s) + \Gamma(\boldsymbol{q}s^{'})}{2}
\end{aligned}
\end{equation}
\end{widetext}
In this equation, the diagonal terms (where $s = s^{'}$) correspond to the $\kappa^{\mathrm{P}}$, while the off-diagonal terms (where $s \ne s^{'}$) account for the $\kappa^{\mathrm{C}}$.
Here, $\Gamma(\boldsymbol{q}s)$ denotes the linewidth of phonon mode ($\boldsymbol{q}s$), defined as $\Gamma(\boldsymbol{q}s) = \frac{1}{2\tau}$; $\hbar$ is the reduced Planck constant; $V$ is the unit cell volume; $N$ is the number of q-points sampled in the Brillouin zone; $\omega(\boldsymbol{q}s)$ is the phonon angular frequency; $n(\boldsymbol{q}s)$ is the BE distribution at temperature $T$; and $v_{\boldsymbol{q}, ss^{'}}^{\alpha}$ represents the $\alpha$th component of the velocity operator with the eigenstates of phonon modes $\boldsymbol{q}s$ and $\boldsymbol{q}s^{'}$~\cite{WignerTransport_PRX_2022}.
\par
For completeness, below we briefly summarize the Green-Kubo formalism for calculating the thermal conductivity from equilibrium MD simulations. 
Within the framework of Onsager's linear response theory, the induced heat flux ($\mathbf{J}$) is assumed to respond linearly to an infinitesimal temperature gradient ($\nabla T$) generated by an external perturbation: $\mathbf{J} = - \kappa \nabla T$.
Based on Onsager's linear response theory, Green and Kubo demonstrated that the transport coefficient, $\kappa$, can be systematically computed through integrals of time correlation functions of the heat flux~\cite{Green1953_JCP, Kubo1957}:
\begin{equation}\label{eq:GK_Intgrl_TC}
\begin{aligned}
\kappa_{\alpha \beta} = \frac{\Omega}{k_{B}T^{2}} \int_{0}^{\infty} \left< {J_{\alpha}}(t) \cdot {J_{\beta}}(0)\right> \,dt 
\end{aligned}
\end{equation}
Here, $\Omega$ denotes the total system volume, and the heat flux, $\mathbf{J}$, is computed from the per-atom energies and stress tensors obtained from MLIP-based MD simulations, thereby incorporating the full many-body nature of the interatomic interactions.
The GK results presented here are adapted from our previous work~\cite{sml_2026}; readers are referred to the corresponding reference for a detailed description of the methodology and computational protocol.

\section{\label{sec:Results} Results and Discussion}
\subsection{\label{subsec:TwoChannelTransport}Phonon Transport: Populations' and Coherences' Contribution in \texorpdfstring{$\kappa$}{kappa}}
\begin{figure*}[ht!]
    \includegraphics[width=16.85cm]{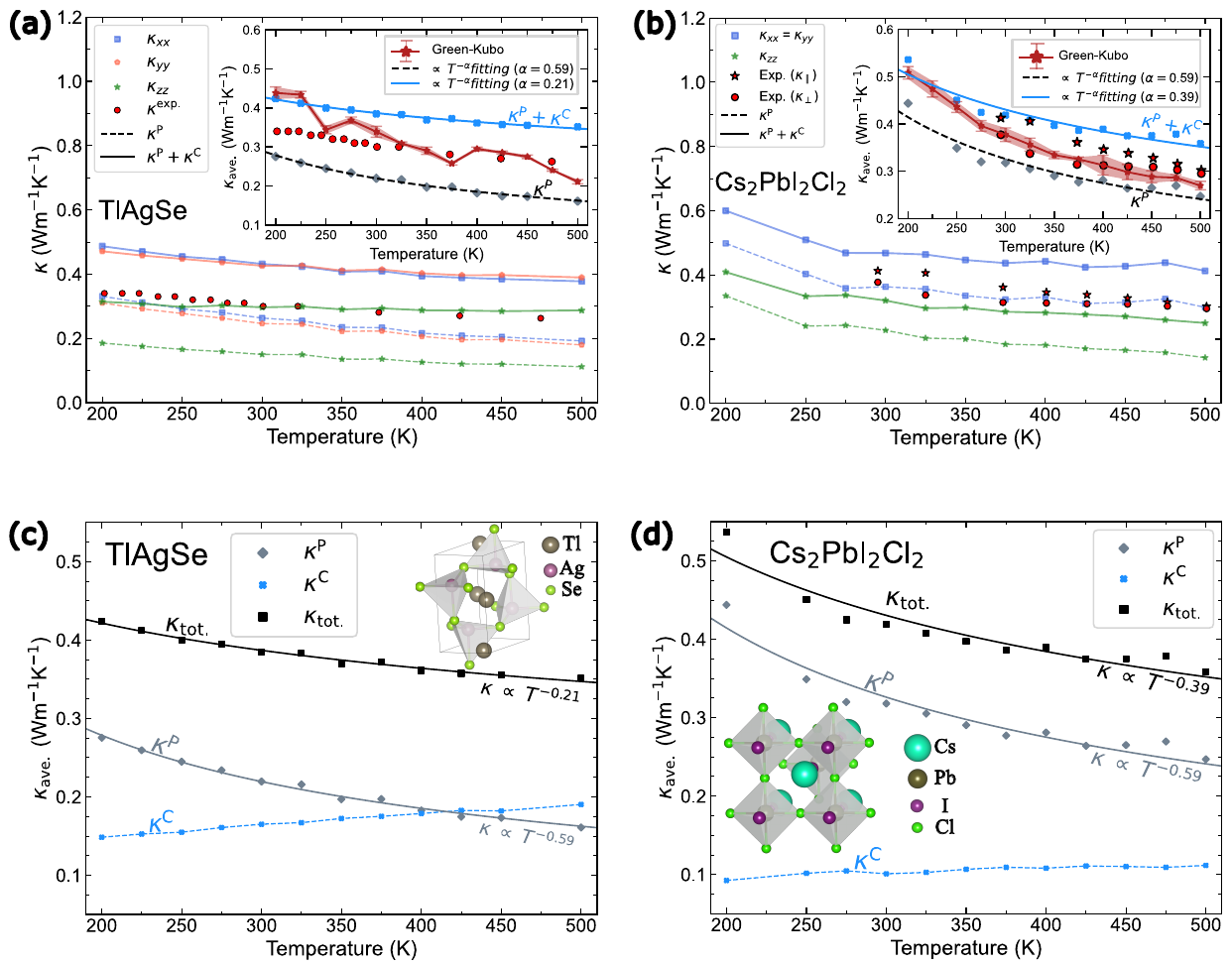}
    \caption[Comparison of thermal conductivity: BTE vs. WTE vs. GK.]{Comparison of thermal conductivity: BTE vs. WTE vs. GK.
(a) Thermal conductivity of TlAgSe predicted by BTE (dashed line) and WTE (solid line) over the 200-500 K range. The coherences' contribution ($\kappa^{\mathrm{C}}$) in WTE leads to an overestimation of $\kappa$ compared to the experiment\cite{TlAgSe_exp}, while the populations' contribution ($\kappa^{\mathrm{P}}$) from BTE underestimates it, as shown in the inset. Power-law fitting yields an $\alpha$ of 0.59 for $\kappa^{\mathrm{P}}_{\mathrm{ave.}}$ (BTE), which decreases to 0.21 with the inclusion of $\kappa^{\mathrm{C}}_{\mathrm{ave.}}$ (WTE), indicating very weak temperature dependence.
The data of $\kappa$ from the Green-Kubo method~\cite{sml_2026} is also presented in the inset plot (red line).
(b) Thermal conductivity of Cs\textsubscript{2}PbI\textsubscript{2}Cl\textsubscript{2} predicted by BTE (dashed line) and WTE (solid line). The $\kappa^{\mathrm{GK}}$ calculated from the Green-Kubo framework~\cite{sml_2026} is presented in the inset plot for comparison. The average $\kappa^{\mathrm{P}}$ slightly underestimates the experimental values\cite{Cs2PbI2Cl2_k-2020}, while including $\kappa^{\mathrm{C}}$ leads to a slight overestimation. Adding the coherences' contribution lowers the $\alpha$ value from 0.59 ($\kappa^{\mathrm{P}}_{\mathrm{ave.}}$) to 0.39, a smaller reduction than seen in TlAgSe, indicating a weaker $\kappa^{\mathrm{C}}$ contribution in Cs\textsubscript{2}PbI\textsubscript{2}Cl\textsubscript{2}.
(c) and (d) Decomposition of the total $\kappa$ into $\kappa^{\mathrm{C}}$ and $\kappa^{\mathrm{P}}$ contributions for TlAgSe (c) and Cs\textsubscript{2}PbI\textsubscript{2}Cl\textsubscript{2} (d). $\kappa^{\mathrm{C}}$ exhibits a glass-like increase with temperature, whereas $\kappa^{\mathrm{P}}$ decreases with increasing temperature. Notably, the relative contribution of $\kappa^{\mathrm{C}}$ is more substantial in TlAgSe than in Cs\textsubscript{2}PbI\textsubscript{2}Cl\textsubscript{2}, and becomes dominant at higher temperatures.
The inset of (c) illustrates the crystal structure of TlAgSe, highlighting a network of interconnected [AgSe\textsubscript{4}]\textsuperscript{-} tetrahedra, with Tl\textsuperscript{+} ions (gray spheres) occupying the interstitial sites. The inset of (d) presents the crystal structure of Cs\textsubscript{2}PbI\textsubscript{2}Cl\textsubscript{2}, composed of [PbI\textsubscript{2}Cl\textsubscript{4}]\textsuperscript{4-} octahedra, with Pb (gray spheres) at the centers and I (red spheres) and Cl (green spheres) at the vertices, linking adjacent octahedral units. Cs\textsuperscript{+} ions (light sky-blue sphere) reside in the interstitial voids, contributing to structural stability.
}
    \label{fig:BTE_WTE_GK}
\end{figure*}
We start by discussing the calculated thermal conductivity from the standard phonon-quasiparticle theory arising from BTE ($\kappa^{\mathrm{BTE}}$) and particle-like propagation and the wave-like coherence dual mode transport arising from WTE ($\kappa^{\mathrm{WTE}}$).
The $\kappa^{\mathrm{WTE}}$ comprises two components: the diagonal term $\kappa^{\mathrm{P}}$, which is equivalent to the phonon-quasiparticle propagation picture in BTE (i.e., $\kappa^{\mathrm{P}} = \kappa^{\mathrm{BTE}}$), and the off-diagonal term ($\kappa^{\mathrm{C}}$), which captures the coherences' contribution associated with diffusive energy transfer between quasi-degenerate eigenstates (i.e., $\kappa^{\mathrm{WTE}} = \kappa^{\mathrm{P}} + \kappa^{\mathrm{C}}$).
It is important to note that the calculation of IFCs follows our in-house-implemented TDEP method\cite{arXiv}, where the force-displacement datasets from the MLIP-based MD trajectory are fitted to calculate IFCs up to the fourth order.
As a result, the IFCs account for all finite-temperature effects consistent with the canonical ensemble, in contrast to the finite displacement (FD) method, which approximates IFCs at 0 K~\cite{ShengBTE_14, Phono3py_FD}.
The incorporation of finite-temperature effects is essential for accurately describing the strongly anharmonic crystalline materials investigated herein, which exhibit temperature-induced structural distortions and local symmetry breaking~\cite{Cs2PbI2Cl2-2024, sml_2026}.
However, the phonon scattering rates in $\kappa^{\mathrm{BTE}}$ and $\kappa^{\mathrm{WTE}}$ for the full range of temperature studied here are calculated considering only up to three-phonon interaction, thereby neglecting higher-order four-phonon scattering processes. 
To assess the role of such higher-order anharmonicity in thermal transport and its impact on agreement with experimental data, four-phonon interactions are computed only at a representative temperature of 300 K, as discussed in the subsequent section. 
The determination of four-phonon scattering matrix elements across all temperatures on the same $\mathbf{q}$-point grid used for the three-phonon calculations is computationally prohibitive for these structurally complex, low-symmetry materials with multiple basis atoms per unit cell.
\par
In Fig.~\ref{fig:BTE_WTE_GK}(a), we have plotted the calculated $\kappa^{\mathrm{BTE}}$ and $\kappa^{\mathrm{WTE}}$ for TlAgSe in the temperature range of 200-500 K.
Our calculations reveal an anisotropic heat-transport coefficient for TlAgSe along the three Cartesian directions.
At 300 K, the computed thermal conductivity for TlAgSe is $\kappa^{\mathrm{BTE}}_{xx}$ $\sim$ 0.26 Wm\textsuperscript{-1}K\textsuperscript{-1}, $\kappa^{\mathrm{BTE}}_{yy}$ $\sim$ 0.24 Wm\textsuperscript{-1}K\textsuperscript{-1}, and $\kappa^{\mathrm{BTE}}_{zz}$ $\sim$ 0.15 Wm\textsuperscript{-1}K\textsuperscript{-1}, yielding an average $\kappa^{\mathrm{BTE}}_{\mathrm{ave.}}$ of $\sim$ 0.22 Wm\textsuperscript{-1}K\textsuperscript{-1}, which underestimates the experimental value of $\sim$ 0.3 Wm\textsuperscript{-1}K\textsuperscript{-1}.\cite{TlAgSe_exp}
As evident from Eq.~\ref{eq:kappaP_C}, the phonon group velocity $v_{\mathbf{q}s}$ governs the directional dependence of $\kappa$ tensor. 
\begin{table*}[!ht]
\setlength\extrarowheight{2.5pt} 
\begin{ruledtabular}
{\renewcommand{\arraystretch}{1.6}
\begin{tabular}{|c|c|c|c|c|c|}
Material & Temperature (K) & $\kappa^{\mathrm{BTE}}$ (Wm\textsuperscript{-1}K\textsuperscript{-1}) & $\kappa^{\mathrm{WTE}}$ (Wm\textsuperscript{-1}K\textsuperscript{-1}) & $\kappa^{\mathrm{GK}}$   (Wm\textsuperscript{-1}K\textsuperscript{-1}) & $\kappa^{\mathrm{exp.}}$ (Wm\textsuperscript{-1}K\textsuperscript{-1}) \\\hline
\multirow{3}{5em}{TlAgSe} & 200K & 0.27 & 0.42 & 0.43 & $\sim$ 0.34\\
\cline{2-6}
& 300K & 0.22 \textbf{(0.17)} & 0.38 \textbf{(0.31)} & 0.37 & $\sim$ 0.30 \\
\cline{2-6}
& 475K & 0.17 & 0.34 & 0.24 & $\sim$ 0.26 \\
\cline{2-6}
\hline
\multirow{3}{5em}{Cs\textsubscript{2}PbI\textsubscript{2}Cl\textsubscript{2}} & 300K & \makecell[l]{$\kappa_{xx}$=$\kappa_{yy}$=0.36 \textbf{(0.30)} \\ $\kappa_{zz}$=0.23 \textbf{(0.20)}} & \makecell[l]{$\kappa_{xx}$=$\kappa_{yy}$=0.47 \textbf{(0.43)}\\ $\kappa_{zz}$=0.32 \textbf{(0.30)}} & \makecell[l]{$\kappa_{xx}$ = 0.46 $\kappa_{yy}$ = 0.43 \\ $\kappa_{zz}$ = 0.24} &  \makecell[l]{$\kappa_{\parallel}$ $\sim$ 0.42 \\ $\kappa_{\perp}$ $\sim$ 0.37} \\
\cline{2-6}
& 400K & \makecell[l]{$\kappa_{xx}$=$\kappa_{yy}$=0.33 \\ $\kappa_{zz}$=0.18 } & \makecell[l]{$\kappa_{xx}$=$\kappa_{yy}$=0.44 \\ $\kappa_{zz}$=0.28 } & \makecell[l]{$\kappa_{xx}$ = 0.40 $\kappa_{yy}$ = 0.32 \\ $\kappa_{zz}$ = 0.22} & \makecell[l]{$\kappa_{\parallel}$ $\sim$ 0.35 \\ $\kappa_{\perp}$ $\sim$ 0.31} \\
\cline{2-6}
& 500K & \makecell[l]{$\kappa_{xx}$=$\kappa_{yy}$=0.30 \\ $\kappa_{zz}$=0.14} & \makecell[l]{$\kappa_{xx}$=$\kappa_{yy}$=0.41 \\ $\kappa_{zz}$=0.25} & \makecell[l]{$\kappa_{xx}$ = 0.36 $\kappa_{yy}$ = 0.26 \\ $\kappa_{zz}$ = 0.18} & \makecell[l]{$\kappa_{\parallel}$ $\sim$ 0.30 \\ $\kappa_{\perp}$ $\sim$ 0.29} \\
\end{tabular}
}
\caption{Table summarizing the calculated thermal conductivity values from the BTE, WTE, and GK frameworks~\cite{sml_2026}. Experimental data are taken from Ref.~\cite{TlAgSe_exp} for TlAgSe and Ref.~\cite{Cs2PbI2Cl2_k-2020} for Cs\textsubscript{2}PbI\textsubscript{2}Cl\textsubscript{2}. In the experimental $\kappa$ for Cs\textsubscript{2}PbI\textsubscript{2}Cl\textsubscript{2}, the $\parallel$ and $\perp$ symbols indicate parallel and perpendicular directions relative to the crystal growth axis. The $\kappa$ values reported for BTE, WTE, and GK represent averages over the three Cartesian directions unless specified otherwise. Values highlighted in bold within parentheses represent $\kappa$ derived from four-phonon scattering calculations.}
\label{tab:Tab1}
\end{ruledtabular}
\end{table*}
Our analysis reveals a significant disparity in the magnitude of $v_{\mathbf{q}s}$ for TlAgSe along the three Cartesian directions, with mode-averaged values of $|v^x|\,\sim$216 ms\textsuperscript{-1}, $|v^y|\,\sim$167 ms\textsuperscript{-1}, and $|v^z|\,\sim$156 ms\textsuperscript{-1}, thereby giving rise to anisotropic transport.
Although the experiments were performed on a single-crystalline TlAgSe sample, the crystallographic orientation of the measurement is not specified. 
Accordingly, for TlAgSe, we compare the orientation-averaged thermal conductivity, $\kappa_{\mathrm{ave.}}$, with the experimental data.
Such underestimation from the phonon-quasiparticle-based BTE model is consistent with the other reports on ultralow-$\kappa$ complex crystals, such as all the inorganic halide perovskites in Ref.\cite{CsBX3_pnas_2017}, and Tl\textsubscript{3}VSe\textsubscript{4} in Ref.\cite{TwoChannelTransport_science_2018}
Accounting for coherence-channel heat conduction in the WTE model gives $\kappa^{\mathrm{WTE}}_{xx}$ $\sim$ 0.43 Wm\textsuperscript{-1}K\textsuperscript{-1}, $\kappa^{\mathrm{WTE}}_{yy}$ $\sim$ 0.42 Wm\textsuperscript{-1}K\textsuperscript{-1}, and $\kappa^{\mathrm{WTE}}_{zz}$ $\sim$ 0.30 Wm\textsuperscript{-1}K\textsuperscript{-1}, resulting in an average $\kappa^{\mathrm{WTE}}_{\mathrm{ave.}}$ of $\sim$ 0.38 Wm\textsuperscript{-1}K\textsuperscript{-1} at 300 K.
It is noteworthy that the contribution of $\kappa^{\mathrm{C}}$ is significant in complex crystals such as TlAgSe.
The $\kappa^{\mathrm{C}}$ contribution dominates as the temperature increases, accounting for $\kappa^{\mathrm{C}}_{\mathrm{ave.}}$ $\sim$ 0.15 Wm\textsuperscript{-1}K\textsuperscript{-1} at 200 K, being 35\% of total $\kappa^{\mathrm{WTE}}_{\mathrm{ave.}}$ ($\kappa^{\mathrm{C}}_{xx}$ $\sim$ 32\%, $\kappa^{\mathrm{C}}_{yy}$ $\sim$ 34\% and $\kappa^{\mathrm{C}}_{zz}$ $\sim$ 41\%) and $\kappa^{\mathrm{C}}_{\mathrm{ave.}}$ $\sim$ 0.19 Wm\textsuperscript{-1}K\textsuperscript{-1} at 500 K, being 54\% of total $\kappa^{\mathrm{WTE}}_{\mathrm{ave.}}$ ($\kappa^{\mathrm{C}}_{xx}$ $\sim$ 49\%, $\kappa^{\mathrm{C}}_{yy}$ $\sim$ 54\% and $\kappa^{\mathrm{C}}_{zz}$ $\sim$ 61\%).
In contrast to $\kappa^{\mathrm{P}}$, which decreases with increasing temperature, $\kappa^{\mathrm{C}}$ exhibits a glass-like rise\cite{Glass_k_Simoncelli2023} as temperature increases.
Power-law fitting of $\kappa^{\mathrm{BTE}}_{\mathrm{ave.}}$ for TlAgSe over the 200–500 K range yields an $\alpha$ value of $\sim$ 0.59, while for $\kappa^{\mathrm{WTE}}_{\mathrm{ave.}}$, $\alpha$ drops sharply to $\sim$ 0.21, as illustrated in the inset of Fig.~\ref{fig:BTE_WTE_GK}(a).
The low $\alpha$ value for $\kappa^{\mathrm{WTE}}$ reflects a very weak temperature dependence, consistent with experimental observations.\cite{TlAgSe_exp}
However, including the coherences' contribution leads to a slight overestimation of $\kappa$ in TlAgSe compared to experimental values\cite{TlAgSe_exp}, as shown in the inset of Fig.~\ref{fig:BTE_WTE_GK}(a), primarily due to neglecting higher-order phonon interactions in the phonon lifetime calculations.
\par
Fig.~\ref{fig:BTE_WTE_GK}(b) shows the calculated $\kappa^{\mathrm{BTE}}$ and $\kappa^{\mathrm{WTE}}$ for Cs\textsubscript{2}PbI\textsubscript{2}Cl\textsubscript{2} across the 200-500 K temperature range. 
As with TlAgSe, incorporating $\kappa^{\mathrm{C}}$ increases the thermal conductivity of Cs\textsubscript{2}PbI\textsubscript{2}Cl\textsubscript{2} relative to the BTE model: at 300 K, $\kappa^{\mathrm{BTE}}_{xx}$ (=$\kappa^{\mathrm{BTE}}_{yy}$) $\sim$ 0.36 Wm\textsuperscript{-1}K\textsuperscript{-1} ($\kappa^{\mathrm{BTE}}_{zz}$ $\sim$ 0.23 Wm\textsuperscript{-1}K\textsuperscript{-1}) compared to $\kappa^{\mathrm{WTE}}_{xx}$ (=$\kappa^{\mathrm{WTE}}_{yy}$) $\sim$ 0.47 Wm\textsuperscript{-1}K\textsuperscript{-1} ($\kappa^{\mathrm{WTE}}_{zz}$ $\sim$ 0.32 Wm\textsuperscript{-1}K\textsuperscript{-1}). 
The average $\kappa^{\mathrm{WTE}}_{\mathrm{ave.}}$ is $\sim$ 0.40 Wm\textsuperscript{-1}K\textsuperscript{-1}, about 38\% higher than $\kappa^{\mathrm{BTE}}_{\mathrm{ave.}}$ ($\sim$ 0.29 Wm\textsuperscript{-1}K\textsuperscript{-1}), at 300 K and comparable to experimental values~\cite{Cs2PbI2Cl2_k-2020} ($\kappa^{\mathrm{exp.}}_{\parallel}$ $\sim$ 0.42 Wm\textsuperscript{-1}K\textsuperscript{-1} and $\kappa^{\mathrm{exp.}}_{\perp}$ $\sim$ 0.37 Wm\textsuperscript{-1}K\textsuperscript{-1}, with $\parallel$ and $\perp$ symbols indicating parallel and perpendicular directions relative to the crystal growth axis).
Analogous to TlAgSe, the anisotropic thermal transport in Cs\textsubscript{2}PbI\textsubscript{2}Cl\textsubscript{2} originates from its structurally anisotropic, layered crystalline structure, together with disparities in the phonon group velocities along the Cartesian directions, with $|v^x| = |v^y|\,\sim$435 ms\textsuperscript{-1} and $|v^z|\,\sim$158 ms\textsuperscript{-1}.
Notably, the coherences' contribution to heat conduction is significantly lower in Cs\textsubscript{2}PbI\textsubscript{2}Cl\textsubscript{2} compared to TlAgSe: at 300 K, $\kappa^{\mathrm{C}}_{\mathrm{ave.}}$ accounts for 42\% of $\kappa^{\mathrm{WTE}}_{\mathrm{ave.}}$ in TlAgSe, whereas in Cs\textsubscript{2}PbI\textsubscript{2}Cl\textsubscript{2} it contributes only 27\%. 
This lower $\kappa^{\mathrm{C}}$ contribution in Cs\textsubscript{2}PbI\textsubscript{2}Cl\textsubscript{2} remains consistent across the entire temperature range.
As with TlAgSe, the $\alpha$ parameter for $\kappa^{\mathrm{BTE}}_{\mathrm{ave.}}$ of Cs\textsubscript{2}PbI\textsubscript{2}Cl\textsubscript{2} is 0.59, as shown in the inset plot of Fig.~\ref{fig:BTE_WTE_GK}(b).
However, incorporating $\kappa^{\mathrm{C}}$ within the Wigner framework reduces the $\alpha$ value to 0.39 for Cs\textsubscript{2}PbI\textsubscript{2}Cl\textsubscript{2}, considerably higher than the $\sim$ 0.21 observed for TlAgSe, indicating a weaker coherences' contribution in Cs\textsubscript{2}PbI\textsubscript{2}Cl\textsubscript{2}.
\par
The temperature dependence of the respective contributions from $\kappa_{\mathrm{ave.}}^{\mathrm{P}}$ and $\kappa_{\mathrm{ave.}}^{\mathrm{C}}$ to the total thermal conductivity, $\kappa_{\mathrm{tot.}}$ (or $\kappa^{\mathrm{WTE}}$), is presented in Fig.~\ref{fig:BTE_WTE_GK}(c) and (d) for TlAgSe and Cs\textsubscript{2}PbI\textsubscript{2}Cl\textsubscript{2}, respectively. For TlAgSe, $\kappa_{\mathrm{ave.}}^{\mathrm{P}}$ constitutes the dominant contribution to $\kappa_{\mathrm{tot.}}$ in the low-temperature regime.
With increasing temperature, however, $\kappa_{\mathrm{ave.}}^{\mathrm{C}}$ rises and ultimately surpasses $\kappa_{\mathrm{ave.}}^{\mathrm{P}}$, with a crossover occurring at approximately 425 K.
\begin{figure*}
    \includegraphics[width=0.82\textwidth]{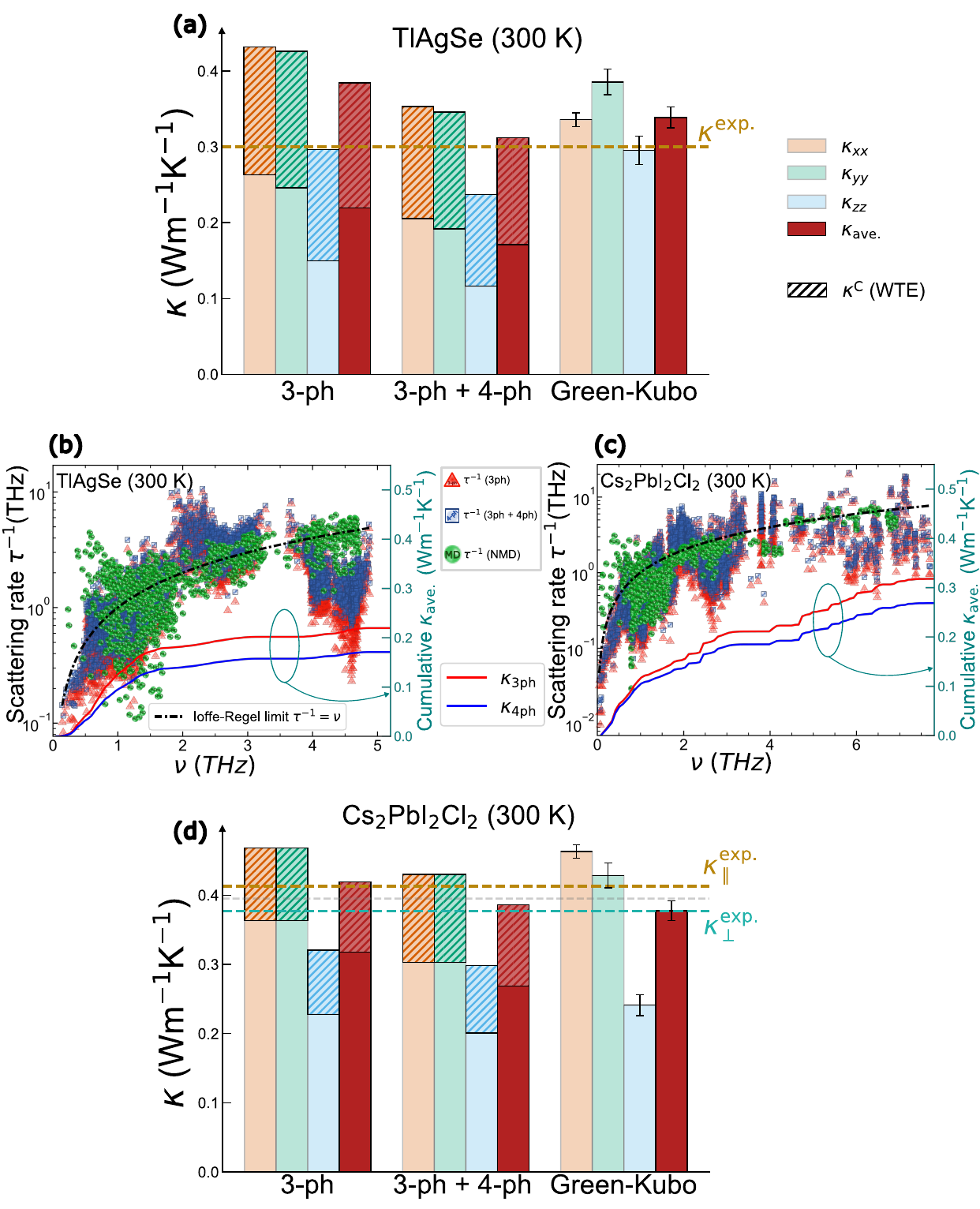}
    \caption[Impact of four-phonon scattering on thermal transport]{Impact of four-phonon scattering on thermal transport. (a) and (d) Bar plots of the calculated $\kappa$ at 300 K obtained from different theoretical frameworks of phonon transport in solids for TlAgSe (a) and Cs\textsubscript{2}PbI\textsubscript{2}Cl\textsubscript{2} (d). Solid bars denote the populations' contribution ($\kappa^{\mathrm{P}}$) arising from the phonon quasiparticle picture within the BTE framework, while the regions with slanted lines superimposed on the bars represent the coherences' contribution ($\kappa^{\mathrm{C}}$) from the WTE. The $\kappa$ values obtained from the Green-Kubo method are also included for comparison with the perturbative approaches discussed here. The horizontal lines indicate the experimental $\kappa$ at 300 K for both materials~\cite{TlAgSe_exp, Cs2PbI2Cl2_k-2020}. Notably, inclusion of four-phonon scattering in conjunction with coherences' contributions yields close agreement with experiment, whereas the three-phonon case overestimates $\kappa$.
    (b) and (c) Phonon scattering rates at 300 K calculated using Fermi's golden rule within the perturbative framework for TlAgSe (b) and Cs\textsubscript{2}PbI\textsubscript{2}Cl\textsubscript{2} (c). 
Inclusion of four-phonon interactions (blue cubes) leads to a substantial increase in the total scattering rates across all phonon modes compared to the three-phonon case (red triangles), resulting in a significant reduction of $\kappa^{\mathrm{P}}$. This reduction in $\kappa^{\mathrm{P}}$ is also reflected in the cumulative $\kappa^{\mathrm{P}}$ as a function of phonon frequency plot. The scattering rates calculated from the nonperturbative NMD method~\cite{sml_2026, NMD_McGaugheyJAP_2013}, directly using the MD trajectory, are also presented (green circle) for comparison with the perturbative calculations.}
    \label{fig:BTE_WTE-4ph}
\end{figure*}
In contrast, for Cs\textsubscript{2}PbI\textsubscript{2}Cl\textsubscript{2}, $\kappa_{\mathrm{ave.}}^{\mathrm{C}}$ remains consistently lower than $\kappa_{\mathrm{ave.}}^{\mathrm{P}}$ over the entire temperature range of 200-500 K, contributing as a minority component to $\kappa_{\mathrm{tot.}}$ with a fractional share not exceeding $\sim$30\%.
This behavior can be attributed to the more complex crystal structure of TlAgSe relative to Cs\textsubscript{2}PbI\textsubscript{2}Cl\textsubscript{2}, which gives rise to a densely packed phonon band structure wherein the phonon linewidths approach or exceed the interband spacing, thereby increasing the contribution from coherence-channel transport.
Notably, $\kappa_{\mathrm{ave.}}^{\mathrm{C}}$ exhibits a glass-like temperature dependence, increasing with temperature (blue line in Fig.~~\ref{fig:BTE_WTE_GK}(c) and (d)), in clear contrast to $\kappa_{\mathrm{ave.}}^{\mathrm{P}}$, which decreases over the same temperature range (gray line in Fig.~~\ref{fig:BTE_WTE_GK}(c) and (d)).
This behavior of $\kappa_{\mathrm{ave.}}^{\mathrm{C}}$ leads to a comparatively weak temperature dependence of the total $\kappa$.

\subsection{\label{subsec:4-ph}Impact of Higher-Order Anharmonicity: Four-Phonon Scattering}
While the inclusion of coherence-mediated transport channel substantially mitigates the underestimation of $\kappa$ relative to experimental data, it introduces a systematic overestimation of the thermal conductivity.
This discrepancy is more significant for TlAgSe (inset of Fig.~\ref{fig:BTE_WTE_GK}(a)) than for Cs\textsubscript{2}PbI\textsubscript{2}Cl\textsubscript{2} (inset of Fig.~\ref{fig:BTE_WTE_GK}(b)).
This systematic overestimation calls for a careful assessment of the role of higher-order anharmonicity within the perturbative framework of thermal transport in solids.
Moreover, from the analysis of the degree of anharmonicity, it is evident that at 300 K, the majority of the contributions in the atomic forces arise from the anharmonic components ($\sim$56 \% for TlAgSe and $\sim$51 \% for Cs\textsubscript{2}PbI\textsubscript{2}Cl\textsubscript{2})~\cite{sml_2026}.
Consequently, restricting the perturbative treatment to the leading-order cubic anharmonic term is insufficient to account for the experimentally observed ultralow $\kappa$ in these strongly anharmonic crystals.
Fig.~\ref{fig:BTE_WTE-4ph} illustrates the effect of incorporating four-phonon scattering processes, arising from quartic anharmonicity, on the predicted thermal transport coefficients of TlAgSe and Cs\textsubscript{2}PbI\textsubscript{2}Cl\textsubscript{2}.
However, the inclusion of four-phonon scattering alone further suppresses the population contribution to the heat transport coefficient ($\kappa^{\mathrm{P}}(\mathrm{4-ph})$), leading to values that deviate even more from the experimental measurements.
Quantitative agreement with experiment is recovered only when the $\kappa^{\mathrm{C}}$ is incorporated alongside the population channel.
\par
For TlAgSe, the four-phonon calculation yields an average population contribution of $\kappa^{\mathrm{P}}_{\mathrm{ave.}} (\mathrm{4-ph})$ = 0.17 Wm\textsuperscript{-1}K\textsuperscript{-1} ($\kappa^{\mathrm{P}}_{xx} (\mathrm{4-ph})$ = 0.21, $\kappa^{\mathrm{P}}_{yy} (\mathrm{4-ph})$ = 0.19, $\kappa^{\mathrm{P}}_{zz} (\mathrm{4-ph})$ = 0.12), corresponding to an underestimation of approximately 43\% relative to the experimental value of $\sim$ 0.3 Wm\textsuperscript{-1}K\textsuperscript{-1}. 
Similarly, for Cs\textsubscript{2}PbI\textsubscript{2}Cl\textsubscript{2}, $\kappa^{\mathrm{P}}_{\mathrm{ave.}} (\mathrm{4-ph})$ = 0.27 Wm\textsuperscript{-1}K\textsuperscript{-1} ($\kappa^{\mathrm{P}}_{xx} (\mathrm{4-ph})$ = $\kappa^{\mathrm{P}}_{yy} (\mathrm{4-ph})$ = 0.30, $\kappa^{\mathrm{P}}_{zz} (\mathrm{4-ph})$ = 0.20), which significantly underpredicts the experimentally reported values of $\kappa_{\parallel}^{\mathrm{exp.}}$ $\sim$0.42 and $\kappa_{\perp}^{\mathrm{exp.}}$ $\sim$0.37 Wm\textsuperscript{-1}K\textsuperscript{-1} at room temperature.
These discrepancies are effectively resolved upon incorporating the coherence-mediated transport channel within the WTE framework in conjunction with four-phonon scattering. 
The resulting average thermal conductivities are $\kappa^{\mathrm{WTE}}_{\mathrm{ave.}} (\mathrm{4-ph})$ = 0.31 Wm\textsuperscript{-1}K\textsuperscript{-1} ($\kappa^{\mathrm{WTE}}_{xx} (\mathrm{4-ph})$ = 0.35, $\kappa^{\mathrm{WTE}}_{yy} (\mathrm{4-ph})$ = 0.34, $\kappa^{\mathrm{WTE}}_{zz} (\mathrm{4-ph})$ = 0.24) for TlAgSe, and $\kappa^{\mathrm{WTE}}_{\mathrm{ave.}} (\mathrm{4-ph})$ = 0.38 Wm\textsuperscript{-1}K\textsuperscript{-1} ($\kappa^{\mathrm{WTE}}_{xx} (\mathrm{4-ph})$ = $\kappa^{\mathrm{WTE}}_{yy} (\mathrm{4-ph})$ = 0.43, $\kappa^{\mathrm{WTE}}_{zz} (\mathrm{4-ph})$ = 0.29) for Cs\textsubscript{2}PbI\textsubscript{2}Cl\textsubscript{2}, both of which are in excellent agreement with the experimental data.
A comparative analysis with three-phonon and four-phonon $\kappa$ values, as well as the respective contribution of $\kappa^{\mathrm{P}}$ and $\kappa^{\mathrm{C}}$, and their correspondence with experimental measurements~\cite{TlAgSe_exp, Cs2PbI2Cl2_k-2020} and Green-Kubo results~\cite{sml_2026} at 300 K, is presented in Fig.~\ref{fig:BTE_WTE-4ph}(a) for TlAgSe and Fig.~\ref{fig:BTE_WTE-4ph}(d) for Cs\textsubscript{2}PbI\textsubscript{2}Cl\textsubscript{2}.
\par
The phonon scattering rates from the three-phonon and four-phonon scattering processes calculated from Fermi's golden rule are shown in Fig.~\ref{fig:BTE_WTE-4ph}(b) for TlAgSe and Fig.~\ref{fig:BTE_WTE-4ph}(c) for Cs\textsubscript{2}PbI\textsubscript{2}Cl\textsubscript{2}, respectively.
The inclusion of four-phonon scattering leads to an overall enhancement of the phonon scattering rates ($\tau^{-1}$) relative to the three-phonon-only case. 
This corresponding increase in $\tau^{-1}$ results in a reduction of the thermal conductivity. 
The consequent suppression of ($\kappa^{\mathrm{P}}_{\mathrm{ave.}}$), as obtained within the BTE framework, is further validated by the cumulative thermal conductivity as a function of the frequency of phonon modes, as illustrated in Fig.~\ref{fig:BTE_WTE-4ph}(b) and Fig.~\ref{fig:BTE_WTE-4ph}(c).
The $\kappa^{\mathrm{P}}_{\mathrm{ave.}} (\mathrm{4-ph})$ obtained from the four-phonon calculations (blue curve) is consistently lower across the entire frequency spectrum compared to the three-phonon case (red curve).
We note that the perturbative calculations of $\kappa$ presented here also explicitly include contributions from phonon-isotopic impurity scattering.
For comparison, we also report $\tau^{-1} (\mathrm{NMD})$ calculated via the normal mode decomposition (NMD) method~\cite{NMD_McGaugheyJAP_2013, sml_2026}, extracted directly from the MD force and displacement data, as indicated by the green circle markers in Fig.~\ref{fig:BTE_WTE-4ph}(c) for TlAgSe and Fig.~\ref{fig:BTE_WTE-4ph}(d) for Cs\textsubscript{2}PbI\textsubscript{2}Cl\textsubscript{2}.
It is noteworthy that the nonperturbative $\tau^{-1} (\mathrm{NMD})$ inherently incorporates phonon interactions to all orders, which is reflected in the distinct nature of the $\tau^{-1} (\mathrm{NMD})$ in the plots compared to the perturbative calculations of $\tau^{-1} (\mathrm{3-ph/4-ph})$ from Fermi's golden rule.
Notably, in all cases, $\tau^{-1}$ for a significant number of phonon modes exceeds the Ioffe-Regel limit (black curve), thereby calling into question the validity of the phonon quasiparticle picture in such strongly anharmonic systems.
The detailed analysis of the applicability of different phonon transport frameworks, based on the magnitude of the scattering rates, is presented in the following section.
Our results for the room-temperature thermal conductivity of TlAgSe and Cs\textsubscript{2}PbI\textsubscript{2}Cl\textsubscript{2} therefore indicate that, in addition to coherence-mediated heat transport, higher-order phonon interactions play a critical role in governing the phonon transport mechanism.
\begin{figure*}
    \includegraphics[width=0.93\textwidth]{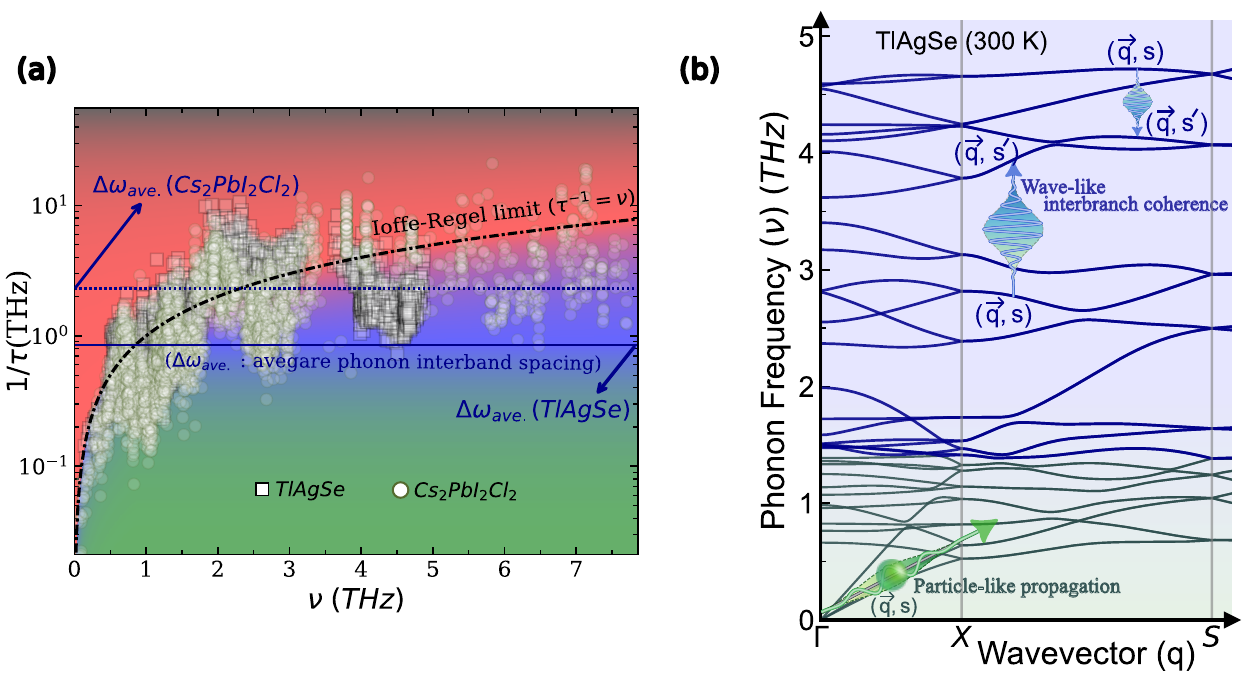}
    \caption{Heat transport regimes. (a) In the phonon frequency ($\nu$) versus scattering rates ($\tau^{-1}$) plot, the green-shaded area approximately indicates the region where $\tau^{-1}$ is much smaller than $\nu$, where the phonon quasiparticle description within the BTE framework is valid.
The blue-shaded region illustrates the dual-mode phonon heat transport, particle-like propagation, and wave-like interbranch coherence, captured by the Wigner formulation. The solid horizontal line marks the average interband spacing ($\Delta \omega_{ave.}$) for TlAgSe, while the dotted line indicates that for Cs\textsubscript{2}PbI\textsubscript{2}Cl\textsubscript{2}. Phonon modes with $\tau^{-1}$ exceeding these lines predominantly contribute to wave-like transport. Notice that a substantial number of phonon modes fall within this region for both materials, though the contribution is smaller for Cs\textsubscript{2}PbI\textsubscript{2}Cl\textsubscript{2} than for TlAgSe due to its larger $\Delta \omega_{ave.}$.
The red-shaded region indicates the heat transport regime beyond the Ioffe-Regel limit ($\tau^{-1} = \nu$), where strong anharmonicity leads to overdamped phonons. In this regime, perturbative approaches such as BTE or WTE break down, and a nonperturbative method like the GK framework is required to capture the heat transport behavior.
(b) Schematic illustration of the populations' ($\kappa^{\mathrm{P}}$) and coherences' ($\kappa^{\mathrm{C}}$) transport overlaid on the phonon dispersion of TlAgSe at 300 K along the high-symmetry path in the Brillouin zone. 
The particle-like propagation of phonons is governed by the phonon group velocity, and is represented by the slope of the dispersion curve indicated by the green arrow at the phonon mode $\mathbf{q}s$. 
In contrast, the interband coherent transport arises from the off-diagonal elements and is depicted with a blue arrow between two different modes with band indices $s$ and $s^{'}$ at the same wavevector $\mathbf{q}$.}
    \label{fig:transport_regime}
\end{figure*}
\subsection{\label{subsec:TransportRegime}Heat Transport Regimes in Anharmonic Solids}
The strong anharmonicity-induced broad phonon linewidths and densely packed phonon branches in these complex ultralow-$\kappa$ crystals make the contribution of the glass-like interbranch coherence mechanism to heat transport significant for accurately capturing experimental observations. 
This phenomenon has been extensively investigated in previous theoretical studies on various ultralow-$\kappa$ materials~\cite{Cu12Sb4S13_PRL_coh_20, TlInTe2_coh_npj_comp_21, Cs3Bi2I6Cl3_PRB_coh_23, Bi4O4SeCl2_nlett_23, KCu7S4_coh_AdvFuncMat_25, 2DMat_PRB_2025, CsCu4Se3_PRB_coh_25, Cs3Bi2I6Cl3_Cs3Bi2Br_PRB_coh_25, PRXE_dual_channel_SolidElectr_25, OrgSemCond_coh_npj_25, TlAgTe_PRB_2026}, including well-studied examples such as Ag-based argyrodites~\cite{Argyrodites_MatTodPhy_23, Ag-argyrodites_coh_PRB_2025, PRXE_k_dual_chnl_25}, Tl\textsubscript{3}VSe\textsubscript{4}\cite{Tl3VSe4_coh_MLIP_MDlifetime_21, TwoChannelTransport_natcommn_2020, TwoChannelTransport_science_2018, Coherent_k_AJain_2020_Tl3VSe4, Tl2VSe4_PRL_2020_coh_trans}, La\textsubscript{2}Zr\textsubscript{2}O\textsubscript{7}\cite{TwoChannelTransport_natcommn_2020, WignerTransport_PRX_2022}, CsPbBr\textsubscript{3}\cite{TwoChannelTransport_2019, WignerTransport_PRX_2022}, and Cs\textsubscript{2}AgBiBr\textsubscript{6}\cite{Cs2AgBiBr6_2024_npjComp}. 
The $\kappa^{\mathrm{C}}$ in the WTE originates from the coherence between distinct phonon bands at the same wave vector $\boldsymbol{q}$: the superposition of two such bands carries a heat flux through the off-diagonal elements of the momentum operator, giving rise to a diffusive transport channel that has no counterpart in the particle-like BTE picture.
The inter-band coherences' transport mechanism coupling two different phonon modes of the same wavevector but distinct band indices $s$ and $s^{'}$ is schematically illustrated by the blue arrow in Fig.~\ref{fig:transport_regime}(b), alongside the phonon dispersion of TlAgSe evaluated along the high-symmetry Brillouin zone path at 300 K.
Note that the $\kappa^{\mathrm{C}}$ is governed by the off-diagonal matrix elements of the momentum operator ($v_{\boldsymbol{q}, ss^{'}}^{\alpha}$ in Eq.~\ref{eq:kappaP_C}), whereas $\kappa^{\mathrm{P}}$ is determined by the diagonal component, i.e., phonon group velocity ($v_{\boldsymbol{q}s}^{\alpha}$).
The latter is represented by the local slope of the phonon dispersion at a given $\boldsymbol{q}s$, as indicated by the green arrow in Fig.~\ref{fig:transport_regime}(b). 
$\kappa^{\mathrm{C}}$ grows as the coherence length between phonon modes $\boldsymbol{q}s$ and $\boldsymbol{q}s^{'}$, $\ell_{\mathrm{coh}} \sim v/ |\Delta\omega_{ss^{'}}|$,  increases (here, $v$ is the sound velocity in the material and $|\Delta\omega_{ss^{'}}| = |\omega(\boldsymbol{q}s) - \omega(\boldsymbol{q}s^{'})|$).
Coherent transport is effective when $\ell_{\mathrm{coh}}$ exceeds the mean free path $\ell_{\mathrm{mfp}} \sim v/ \Gamma_{ss^{'}}$, i.e., the average line width ($\Gamma_{ss^{'}} = \Gamma(\boldsymbol{q}s) + \Gamma(\boldsymbol{q}s^{'})/2$) crosses the interband spacing ($\Gamma_{ss^{'}} > |\Delta\omega_{ss^{'}}|$).
The structurally complex materials investigated in this work exhibit densely packed phonon bands, leading to small $|\Delta\omega_{ss^{'}}|$, while their strong anharmonicity substantially broadens the phonon linewidths. 
Together, these effects satisfy the condition $\Gamma_{ss^{'}} > |\Delta\omega_{ss^{'}}|$ (see denominator of Eq.~\ref{eq:kappaP_C}), thereby promoting coherence-mediated heat transport.
\par
The Wigner limit~\cite{WignerTransport_PRX_2022}, which defines the crossover from particle-like to wave-like phonon heat conduction, is characterized by $\tau^{-1} > \Delta \omega_{\mathrm{ave.}}$, where $\Delta \omega_{\mathrm{ave.}}$ (=$\frac{\omega_{\mathrm{max.}}}{3N_c}$) represents the average interband spacing, calculated as the ratio of the maximum phonon angular frequency ($\omega_{\mathrm{max.}}$) to the total number of bands ($3N_c$, with $N_c$ being the number of basis atoms per unit cell).
In complex crystals (where $N_c$ is higher) with ultralow thermal conductivity, strong anharmonicity-driven phonon scattering rates ($\tau^{-1}$) can surpass $\Delta \omega_{\mathrm{ave.}}$, as these materials generally have a lower $\omega_{\mathrm{max.}}$.
Fig.~\ref{fig:transport_regime}(a) illustrates a diagrammatic representation of the different transport regimes, along with the phonon scattering rates for TlAgSe and Cs\textsubscript{2}PbI\textsubscript{2}Cl\textsubscript{2} at 300 K. 
The green-shaded area denotes the region where particle-like phonon propagation dominates, while the blue-shaded area roughly indicates the regime where wave-like interbranch coherence prevails.
Notably, a substantial portion of the phonon modes exhibit scattering rates that fall within the wave-like transport regime of the WTE for both TlAgSe and Cs\textsubscript{2}PbI\textsubscript{2}Cl\textsubscript{2}, as highlighted by the blue-shaded region in Fig.~\ref{fig:transport_regime}(a).
However, the Wigner limit is higher in Cs\textsubscript{2}PbI\textsubscript{2}Cl\textsubscript{2} than in TlAgSe, owing to its larger $\omega_{\mathrm{max.}}$ ($\omega_{\mathrm{max.}}$ $\sim$ 30 rad$\cdot$ps\textsuperscript{-1} in TlAgSe compared to 44 rad$\cdot$ps\textsuperscript{-1} in Cs\textsubscript{2}PbI\textsubscript{2}Cl\textsubscript{2}) and a smaller number of basis atoms (7 versus 12 in TlAgSe). 
In Fig.~\ref{fig:transport_regime}(a), the Wigner limit is indicated by a solid and dotted horizontal line for TlAgSe and Cs\textsubscript{2}PbI\textsubscript{2}Cl\textsubscript{2}, respectively.
The higher Wigner limit in Cs\textsubscript{2}PbI\textsubscript{2}Cl\textsubscript{2} reduces the prevalence of coherent phonon conduction, as seen in Fig.~\ref{fig:transport_regime}(a), where significantly fewer phonon modes have $\tau^{-1}$ exceeding the limit compared to TlAgSe.
This is reflected in our calculated $\kappa^{\mathrm{WTE}}$, where $\kappa^{\mathrm{C}}_{\mathrm{ave.}}$ accounts for 27\% of $\kappa^{\mathrm{WTE}}_{\mathrm{ave.}}$ in Cs\textsubscript{2}PbI\textsubscript{2}Cl\textsubscript{2} compared to 42\% in TlAgSe at 300 K, increasing to 31\% and 54\%, respectively, at 500 K.
\par
Another key regime in Fig.~\ref{fig:transport_regime}(a), marked by the red-shaded area, represents highly anharmonic materials where phonons become overdamped. 
In this regime, the phonon quasiparticle picture breaks down, casting doubt on the validity of perturbative approaches such as BTE or WTE, as these materials are not confined to displacements around a single stable equilibrium structure and often exhibit metastable defect configurations~\cite{Instability_perovskite_PRXE_24, AL_anhMat_2025}.
The Ioffe-Regel limit\cite{IoffeRegel} defines the boundary for the validity of well-defined phonon quasiparticles, requiring that their scattering rates ($\tau^{-1}$) remain much below their oscillation frequencies ($\nu$) for the quasiparticle picture to hold (i.e., $\tau^{-1} < \nu$).
Fig.~\ref{fig:transport_regime}(a) illustrates that for both TlAgSe and Cs\textsubscript{2}PbI\textsubscript{2}Cl\textsubscript{2} at 300 K, scattering rates of a substantial number of phonon modes, including those at low frequencies, surpass the Ioffe-Regel limit, indicated by the black dotted line.
Rooted in linear response theory, the nonperturbative GK framework can describe thermal transport in materials with an arbitrary degree of anharmonicity, including those that are beyond the Ioffe-Regel limit.
As such, the GK framework's analysis of thermal conductivity is especially sought after for such strong anharmonic materials. 
Our $\kappa^{\mathrm{GK}}$ predictions show much closer agreement with experimental values for both TlAgSe and Cs\textsubscript{2}PbI\textsubscript{2}Cl\textsubscript{2} compared to the BTE and WTE approaches, as illustrated in the inset of Fig.~\ref{fig:BTE_WTE_GK}(a) and Fig.~\ref{fig:BTE_WTE_GK}(b) and summarized in Table~\ref{tab:Tab1}.
However, as the GK approach is based on the classical MD simulations, it neglects the BE statistics of phonons and nuclear quantum effects. 
Consequently, it systematically overestimates the $\kappa$ value at low temperatures (i.e., below the Debye temperature), as observed for TlAgSe in the present case. 
The inset of Fig.~\ref{fig:BTE_WTE_GK}(a) highlights this overprediction of $\kappa^{\mathrm{GK}}$ relative to experimental measurements at low temperatures, whereas at the higher temperatures, the agreement with experiment is significantly improved.
In contrast, the perturbative BTE/WTE framework explicitly incorporates the full BE statistics. Consequently, when higher-order phonon scattering and coherence-mediated transport channels are included, the perturbative approach achieves agreement with experimental data that is comparable to, or surpasses, that of the nonperturbative GK framework, as demonstrated in Fig.~\ref{fig:BTE_WTE-4ph}(a) for TlAgSe and Fig.~\ref{fig:BTE_WTE-4ph}(d) for Cs\textsubscript{2}PbI\textsubscript{2}Cl\textsubscript{2}.

\section{\label{sec:Conclusions} CONCLUSIONS}
In summary, we have presented a comprehensive comparative analysis of different theoretical heat transport frameworks, evaluated in terms of their accuracy with experimental observations.
Our calculations incorporate full temperature-dependent anharmonic lattice dynamics, enabled by MLIP-based MD simulations, which accurately reproduce the experimentally observed structural and thermal transport properties of TlAgSe and Cs\textsubscript{2}PbI\textsubscript{2}Cl\textsubscript{2}.
Restricting the perturbative treatment to the first order, i.e., three-phonon scattering, proves inadequate for capturing the experimental $\kappa$ of these strongly anharmonic materials. 
Within this approximation, the BTE underestimates $\kappa$, while the WTE yields a slight overestimation relative to both experimental measurements and Green-Kubo predictions.
However, incorporating quartic anharmonicity within the perturbative WTE framework by including computationally expensive four-phonon scattering processes effectively resolves discrepancies with experimental observations. 
This highlights that, in addition to the $\kappa^{\mathrm{C}}$, arising from wave-like interbranch coherence of phonon eigenstates, higher-order anharmonic interactions are also crucial to accurately describe thermal transport in these ultralow-$\kappa$ materials.
The $\kappa^{\mathrm{C}}$ constitutes a significant component of the total thermal conductivity in these strongly anharmonic and complex crystals. 
Its glass-like increase with temperature and eventual dominance over the populations' term ($\kappa^{\mathrm{P}}$) at higher temperatures leads to a weak overall temperature dependence of $\kappa$, in agreement with experimental observations for these materials.
The phonon scattering rates exceed the Wigner limit, indicating a substantial contribution from $\kappa^{\mathrm{C}}$, and also surpass the Ioffe-Regel limit for a significant fraction of phonon modes, thereby challenging the validity of the conventional phonon quasiparticle picture and the BTE-based phonon-gas model of interactions in these materials.
The theoretical and numerical framework developed in this work, based on advanced computational predictive modeling, provides key insights into the fundamental mechanisms governing heat transport in ultralow-$\kappa$ materials, and thereby offers a robust foundation for the rational design of materials for thermoelectric and thermal barrier coating applications.

\section*{\label{sec:Acknowledgements} ACKNOWLEDGMENTS}
The authors thank TUE-CMS, IISc, funded by DST, India, as well as Param-Pravega at SERC, IISc, for providing the computational facilities. S.M. acknowledges the generous fellowship provided by MHRD, India. PKM  acknowledges funding through SERB, IRHPA (No. IPA/2020/000034), and ANRF, India, for support through the JC Bose Grant (ANRF/JBG/2025/000227/PS). 
A.S. acknowledges the DST-INSPIRE fellowship
[IF190068]. A.S. and A.K.S. extend their thanks to the Materials Research Center (MRC) at the Indian Institute of Science, Bangalore, for providing essential computational facilities.
The authors acknowledge Prof. Kanishka Biswas for providing the experimentally synthesized structures of the two materials investigated in this study and for useful discussions.

\section*{\label{sec:DataAvailability} DATA AVAILABILITY}
The MLIPs developed for TlAgSe and Cs\textsubscript{2}PbI\textsubscript{2}Cl\textsubscript{2}, together with the corresponding \texttt{LAMMPS} input scripts, are freely available through the GitHub repository: \url{https://github.com/shm-phy/TlAgSe_Cs2PbI2Cl2_MLIP}. 
The finite-temperature IFCs and thermal conductivity calculations were performed using our in-house developed \texttt{SPRING} code, which is openly accessible at the GitHub repository: \url{https://github.com/shm-phy/SPRING}. 
Additional data supporting the findings of this study are available from the corresponding author upon reasonable request.
\nocite{*}

\bibliography{citations}

\end{document}